\documentclass[dvips]{article}

\usepackage{icrc2011}

\title{The endpoint formalism for the calculation of electromagnetic radiation and its applications in astroparticle physics}
\newcommand{\etal}{\MakeLowercase{\textit{et al. }}} % "et al."
\shorttitle{T. Huege \etal The endpoint formalism for the calculation of electromagnetic radiation}

\authors{T. Huege$^{1}$, C. W. James$^{2}$, H. Falcke$^{2,3}$, M. Ludwig$^{4}$}

\afiliations{$^1$Karlsruher Institut f\"ur Technologie, Institut f\"ur Kernphysik, Campus Nord, 76021 Karlsruhe, Germany\\
$^2$Department of Astrophysics, IMAPP, Radboud University Nijmegen, P.O.Box 9010, 6500GL Nijmegen, The Netherlands\\
$^3$ASTRON, Postbus 2, NL-79900AA Dwingeloo, the Netherlands\\
$^4$Karlsruher Institut f\"ur Technologie, Institut f\"ur Experimentelle Kernphysik, Campus S\"ud, 76128 Karlsruhe, Germany
}

\email{tim.huege@kit.edu}

\abstract{We present the ``endpoint'' formalism for the calculation of electromagnetic radiation and illustrate its applications in astroparticle physics. We use the formalism to explain why the coherent radiation from the Askaryan effect is not in general Cherenkov radiation, as the emission directly results from the time-variation of the net charge in the particle shower. Secondly, we illustrate how the formalism has been
applied in the air shower radio emission code REAS3 to unify the microscopic and macroscopic views of radio emission from extensive air showers. Indeed, the formalism is completely universal and particularly well-suited for implementation in Monte Carlo codes in the time- and frequency-domains. It easily reproduces well-known ``classical mechanisms" such as synchrotron radiation, Vavilov-Cherenkov radiation and transition radiation in the adequate limits, but has the advantage that it continues to work in realistic, complex situations, where the ``classical mechanisms'' tend to no longer apply and adhering to them can result in misleading interpretations.}


\newcommand{\tpo}{t^{\prime}_0}

\begin{document}
\maketitle

%Begin the section.
\section{Introduction}

Processes leading to the emission of electromagnetic radiation can be very complex. To get a grasp of these processes, physicists often revert to ``classical named radiation mechanisms'' such as synchrotron radiation, transition radiation or Vavilov-Cherenkov radiation and try to apply these concepts to the problem at hand. However, it turns out that this strategy can often yield misleading results. Usually, the classical mechanisms only apply to idealized problems involving, e.g., infinite particle tracks, and their application to complex realistic situations can cloud the view for the complete picture.

Here, we propose a formalism suitable for the calculation of electromagnetic radiation from any kind of particle acceleration, which lends itself very well to the implementation in Monte Carlo codes, and necessitates no simplifying approximations to the underlying processes. In this ``endpoint formalism'', presented in detail in Ref.\ \cite{EndpointTheory}, the trajectory of individual particles is described as a series of points at which particles are accelerated instantaneously, leading to discrete radiation contributions which can then be easily superposed.

The endpoint formalism can be applied in both the frequency and time domains, and correctly reproduces the aforementioned classical named processes. Furthermore, we demonstrate how the application of the endpoint formalism has provided essential insights for radio emission processes in astroparticle physics.

\section{The endpoint formalism}\label{formalism}

The endpoint formalism relies on the fact that any trajectory of a charged particle undergoing acceleration can be decomposed into a series of discrete ``acceleration events'' joined by straight tracks. The error introduced by this procedure can be made arbitrarily small by applying the  decomposition on scales as small as necessary.

Furthermore, each of the acceleration events during which the velocity vector of a charged particle changes, can again be decomposed into two ``endpoints'', one corresponding to the instantaneous acceleration of the particle from its velocity to rest and the second corresponding to the instantaneous acceleration of the particle from rest to its new velocity. 

The electromagnetic radiation associated with these endpoints can be calculated with a simple formula, which lends itself well to application in Monte Carlo simulations. (For derivations of the stated equations we kindly refer the reader to \cite{EndpointTheory}.)

\subsection{Frequency domain}

The electric field seen by an observer at position $\vec{x}$ and observing frequency $\nu$ from an individual endpoint is given by:
\begin{eqnarray}
\vec{E}_{\pm}(\vec{x},\nu) & = & \pm \frac{q }{c} \, \frac{e^{i k R(t^{\prime}_0)}}{R(t^{\prime}_0)} \, \frac{e^{2 \pi i \nu \tpo} }{1- n \vec{\beta}^* \cdot \hat{r}} \nonumber \\
& \cdot & \hat{r} \times [\hat{r} \times \vec{\beta}^*] \label{endpoint_eqn_f}
\end{eqnarray}
where $k=2 \pi \nu n/c$, $q$ denotes the particle charge, $n$ is the medium refractive index, $\vec{r}$ gives the line of sight from observer to particle, $R=|\vec{r}|$ and $\tpo$ corresponds to the (retarded) time at which the endpoint radiates. $\vec{\beta^{*}}$ denotes the non-zero velocity associated with the endpoint. `$\pm$' is positive when the acceleration is from rest to $\vec{\beta}^*$ and negative when the acceleration is from $\vec{\beta}^*$ to rest.

\subsection{Time domain}\label{sec:gamma}

Likewise, the electric field an observer sees from an endpoint in the time domain is given by:
\begin{equation}\label{endpoint_eqn_t}
\vec{E}_{\pm}(\vec{x},t) = \pm \frac{1}{\Delta t}\frac{q}{c} \left( \frac{\hat{r} \times [\hat{r} \times \vec{\beta}^*]}{(1 - n \vec{\beta}^* \cdot \hat{r}) R} \right)
\end{equation}
Here, the result has to be interpreted as the electric field time-averaged over a time scale $\Delta t$, the adequate choice of which is dictated by the time resolution of interest.

\subsection{Example application}

\begin{figure}[htb]
\includegraphics[width=0.49\textwidth]{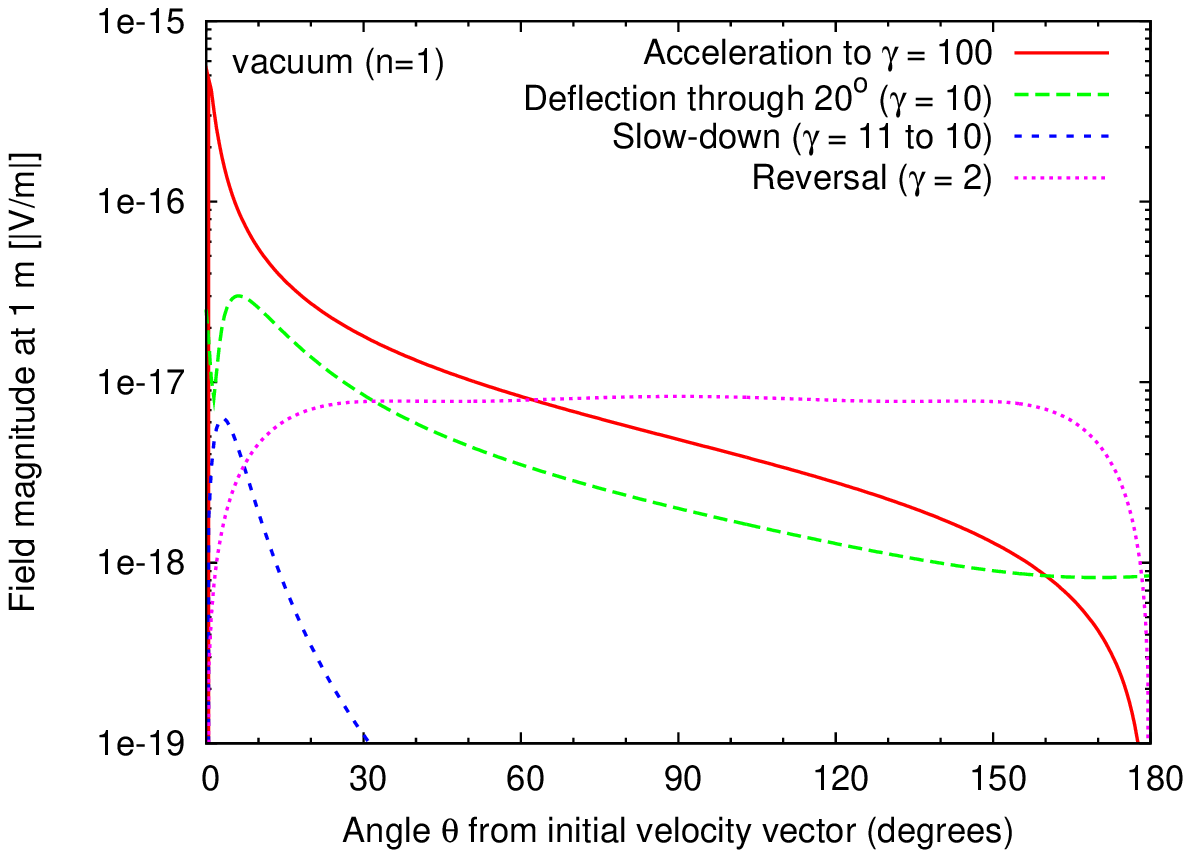}
\includegraphics[width=0.49\textwidth]{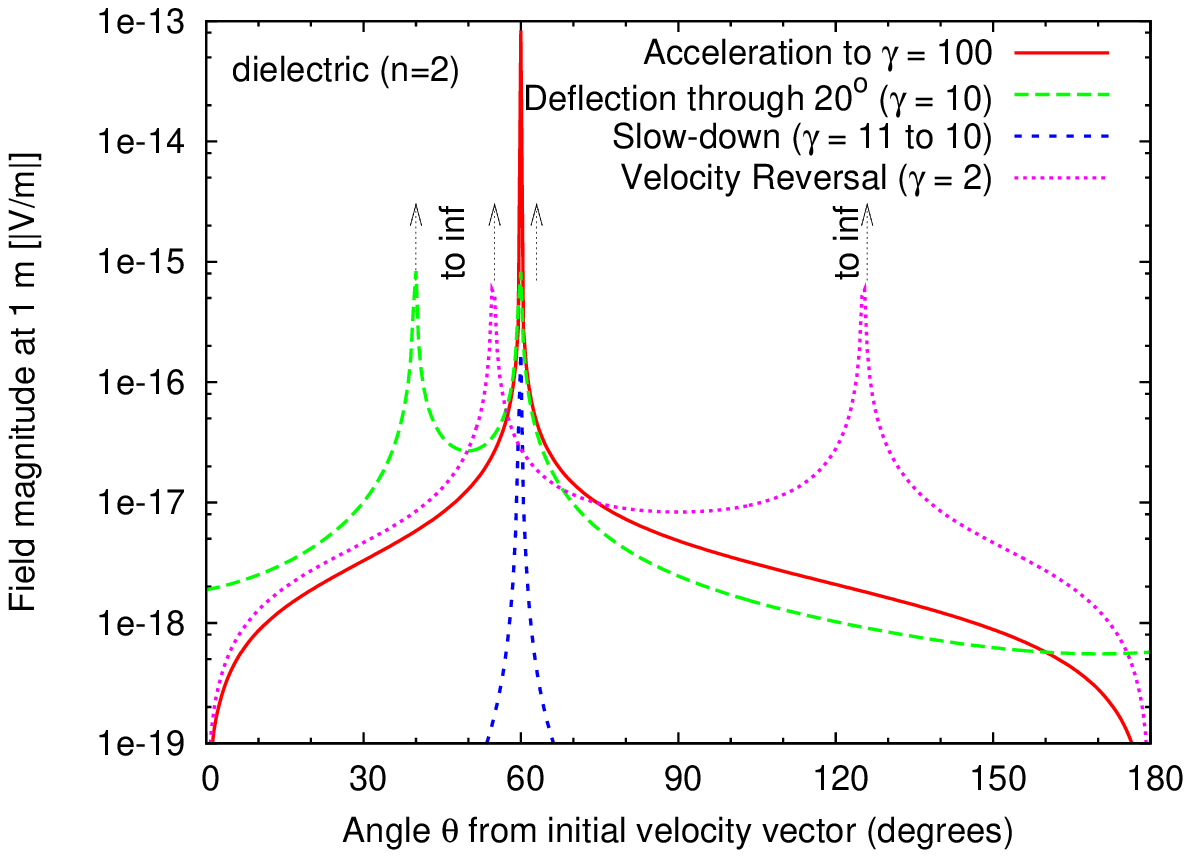}
\caption{Electric field magnitude resulting from the acceleration of a relativistic electron in four simple cases (see text) in (top) vacuum and (bottom) a dielectric with refractive index $n=2$, as calculated using Eq.\ (\ref{endpoint_eqn_f}).} \label{vac_di_fig}
\end{figure}

In Fig.\ \ref{vac_di_fig} we illustrate the electromagnetic radiation associated with the acceleration of individual charged particles as calculated using the endpoint formalism in the frequency domain. In the figure, the emission from processes such as the instantaneous acceleration of a particle from rest (`acceleration'), the deflection of a relativistic electron by 20$^{\circ}$ (`deflection'), the deceleration of a relativistic electron to rest (`slow-down'), and the reversal of direction of a mildly relativistic electron (`reversal'), is shown. Typical effects such as the beaming of radiation for relativistically moving charges and the shock wave behaviour near the Cherenkov angle occurring in media with a refractive index different from unity turn up naturally in this calculation.

\section{Application to classical mechanisms}\label{processes}

To demonstrate that the endpoint formalism can reproduce ``classical named radiation mechanisms'', we briefly illustrate its application to the processes of synchrotron radiation and Vavilov-Cherenkov radiation.

\subsection{Synchrotron radiation}

While the term ``synchrotron radiation'' usually implies relativistic particles revolving continuously in a magnetic field, here we look at a particle undergoing precisely one revolution. To describe this situation using the endpoint formalism, the helical particle track is decomposed into a series of straight tracks joined by ``kinks'' in which the particle is accelerated instantaneously. As explained earlier, each kink in turn is represented by two endpoints with contributions $\vec{E}_{-}$ and $\vec{E}_{+}$, describing the instantaneous deceleration to rest followed by the instantaneous acceleration to the new velocity (in this case with the same magnitude, but different direction).

\begin{figure*}
\begin{center}
\includegraphics[width=0.485\textwidth]{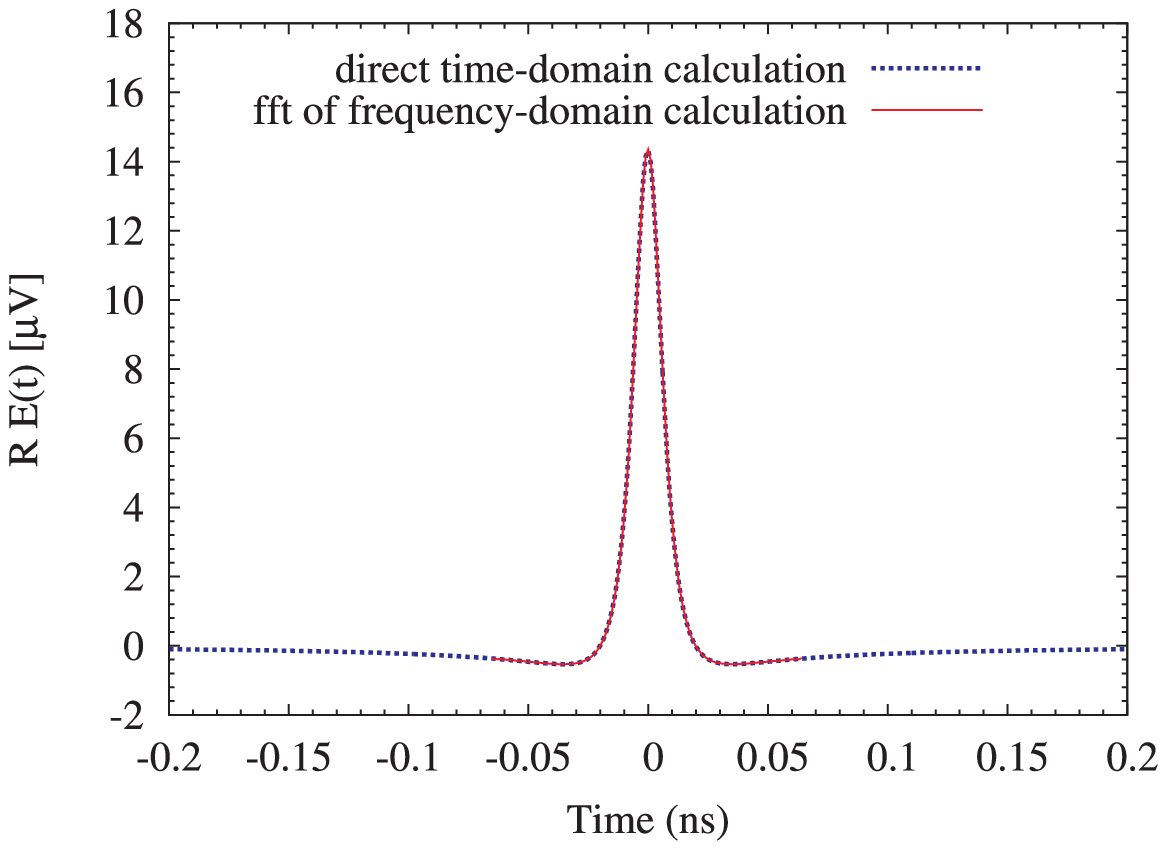}
\includegraphics[width=0.50\textwidth]{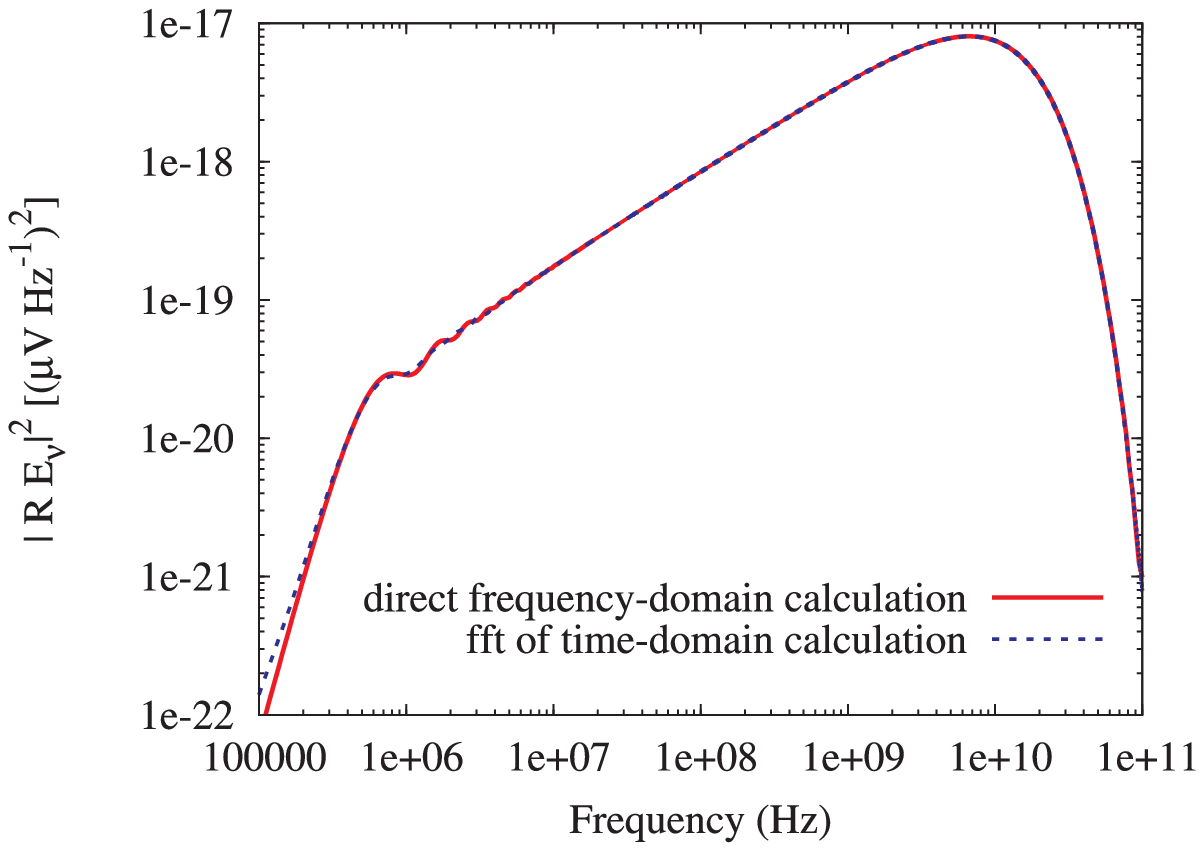}
\end{center}
\caption{Time-trace of a synchrotron pulse produced by a single gyration of an electron with $\beta = 0.999$ and gyration radius $r=100$~m (top), and power frequency spectrum of the emitted radiation (bottom). In each case, the direct calculations (time- and frequency-domain respectively) are shown in comparison with the results generated by fast-Fourier-transforming data from the other domain.\label{synch_fig}}
\end{figure*}

Using either of equations (\ref{endpoint_eqn_f}) and (\ref{endpoint_eqn_t}), the emission radiated during the particle revolution can then be calculated as the superposition of the individual endpoint contributions. Care has to be taken to describe the particle motion on a fine-enough spatial scale, as dictated (respectively) by the maximum frequency/minimum time interval of interest. The results for both the frequency- and time-domain are shown in Fig.\ \ref{synch_fig}. Note that calculating the result for one domain by an FFT of the result of the other domain yields the same result, but is usually much more involved than performing a direct endpoint calculation.

Comparing these results with the established theory of synchrotron radiation confirms that the endpoint formalism can reproduce the well-known theory. (For further details please refer to Ref.\ \cite{EndpointTheory}.)

\subsection{Vavilov-Cherenkov radiation}

In the scientific literature, the term ``(Vavilov)-Cherenkov radiation'' is often used somewhat imprecisely, but typically refers to the original calculation of Frank and Tamm in the case of charged particles propagating through a dielectric medium along infinite tracks \cite{FrankTamm37}. There is no acceleration associated with this particle movement, and therefore this ``true'' Vavilov-Cherenkov emission cannot be described with the endpoint formalism discussed here.

However, in realistic situations particle tracks are not infinite. When trying to calculate the emission associated with particles propagating through a medium on finite tracks, the emission is usually calculated using ``track segments'', and, as first calculated by Tamm \cite{Tamm39}, the far-field result near the Cherenkov angle turns out to be proportionate to the total length of track traversed.

Using the endpoint formalism, we regard the radiation as contributions from the endpoints of the track segments. Calculating the radiation based on equation (\ref{endpoint_eqn_f}), we can derive (cf.\ Ref.\ \cite{EndpointTheory} for details):
\begin{eqnarray}
\vec{E}_{\rm track}(\vec{x}, \nu) & \approx & \frac{q \beta \sin \theta }{c} \frac{e^{i (k R + 2 \pi \nu t_1)}}{R} \nonumber \\*
&& \times \left( \frac{ 1 - e^{2 \pi i \nu (1- n \beta \cos \theta) \delta t}}{1 - n \beta \cos \theta} \right) \hat{E} \label{particle_track}
\end{eqnarray}
where $\theta$ denotes the angle between the particle velocity vector and the line of sight from observer to endpoint, $t_{1}$ denotes the (retarded) time of the start of the track, $\delta t$ is the track length divided by $c$, and an approximation was made for observer distances large with respect to the track length.

This result corresponds (within a factor of $2$ arising from a different definition of the Fourier transform) to the ``Vavilov-Cherenkov radiation formula'' of Eq.\ 12 in Zas, Halzen, and Stanev \cite{ZHS92}, with $\mu_r=1$ and $q=-e$. These authors also show that their result --- and, by extension, ours --- produces the same emitted power as Tamm's calculation. Indeed, it has previously been shown that the radiation from a finite particle track consists of both endpoint contributions {\it and} a ``true'' Vavilov-Cherenkov contribution from the track itself, the latter of which is negligible at most angles in the far-field \cite{Afanasiev99}.

In other words, although the ZHS code is commonly understood as calculating ``the Vavilov-Cherenkov radiation'' from a particle cascade in a dense medium, it actually calculates the radiation from the endpoints of finite particle tracks. As the particle cascade exhibits a variation of the negative charge excess during its evolution, the endpoint contributions of the individual tracks do not cancel and a net contribution arises. This radiation due to the time-variation of the net charge is a central, yet not generally appreciated aspect of the Askaryan radiation \cite{Askaryan} process of cascades in dense media. Thus, we expect that the most recent results on ``\v{C}erenkov radio pulses'' from straight track segments \cite{AMRWZ2010}, wherein the authors use a method based on calculating the vector potential, is generally applicable to all radiation processes.

\section{Applications in astroparticle physics}\label{astro}

We have developed the endpoint formalism in the context of astroparticle physics and hereby illustrate how it has helped us to achieve a better understanding of radio emission processes relevant for cosmic ray and neutrino detection.

\subsection{Radio emission from air showers}

\begin{figure}
\begin{center}
\includegraphics[angle = 270, width=0.485\textwidth]{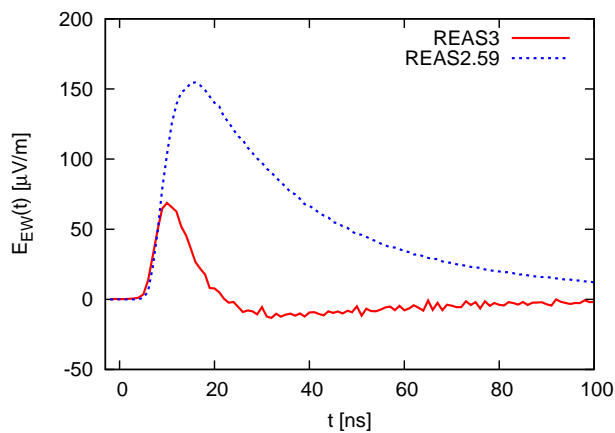}
\end{center}
\caption{A radio pulse simulated with REAS3 (solid red line) compared with the radio pulse simulated with REAS2.59 (blue dashed line). REAS3 is based on the endpoint formalism.\label{reas3_vs_reas2_pulse}}
\end{figure}

Radio emission from extensive air showers has been under intense study over the recent decade. To interpret the data gathered by experiments, a good understanding of the underlying emission mechanism is vital. A significant number of models have been developed in recent years, but different models gave different results. In particular, there was a discrepancy between microscopic Monte Carlo based approaches such as REAS2 \cite{HuegeUlrichEngel2007} and the macroscopic MGMR model \cite{ScholtenWernerRusydi2008}. It became clear that in the microscopic time-domain treatments, a radiation contribution associated with the time-variation of the number of charged particles had been neglected. Using the endpoint formalism in the time-domain in REAS3 \cite{LudwigHuege2010}, it could be understood in detail how the superposition of the individual particle emissions can be carried out consistently. Consequently, the different models could be reconciled \cite{Huege_ARENA} --- a breakthrough in the modelling of radio emission from extensive air showers. The importance of the correction introduced by the endpoint formalism is illustrated in Fig.\ \ref{reas3_vs_reas2_pulse}. 

\subsection{Radio emission in the lunar regolith}

Numerous projects aim at detecting cosmic rays or neutrinos via radio emission radiated from showers initiated in the lunar regolith. For a long time it has been the general understanding that the radio emission would vanish if the particle shower occurred close to the surface of the moon due to the so-called ``formation zone effect'' (first investigated in this context by Ref.\ \cite{Gorham_formation}). The argument is that the refractive index close to the surface effectively approaches unity and that therefore Vavilov-Cherenkov emission has to vanish.

As has been argued above, however, the emission produced by showers in dense media (the Askaryan radiation) is dominated by radiation from the endpoints of individual particle tracks. The fact that the shower grows and declines again means that there is a net radiation contribution due to the variation of the net charge, which leads to radiation irrespective of the refractive index of the medium. A suppression of the emission due to the presence of the surface still occurs due to refractive effects at the surface --- however, this effect applies equally to both shallow and deep cascades, and is taken into account in experimental simulations.

\section{Conclusions}

We have developed an endpoint formalism for the calculation of electromagnetic radiation in arbitrarily complex radiation processes. The formalism is universally applicable and can be implemented easily in calculations both in the time and frequency domains.

Classical named radiation processes such as synchrotron radiation or Vavilov-Cherenkov radiation can be successfully reproduced using the endpoint formalism. However, unlike such descriptions, the endpoint formalism does not make any simplifying assumptions which often break down in realistic situations.

We have successfully used the endpoint formalism to address problems in astroparticle physics. In particular, we were able to calculate the radio emission from extensive air showers using the endpoint formalism, and thereby reconcile discrepancies existing between earlier modelling approaches. Also, the endpoint formalism illustrates that Askaryan radio emission in the lunar regolith does not vanish even if the particle cascade develops close to the surface, as the net growth and decline of the shower will produce radio emission even for a refractive index approaching unity.

%
%\begin{figure}[!t]
%  \vspace{5mm}
%  \centering
%  \includegraphics[angle = 270, width= 0.65\columnwidth]{icrc0149_fig04a.eps}
%  \includegraphics[angle = 270, width= 0.65\columnwidth]{icrc0149_fig04b.eps}
%  \caption{Contribution of different shower evolution stages to the east-west polarization of the raw radio pulses
%		for observers 100\,m (top) and 400\,m (bottom) north of the shower core.}\label{fig:Xdistributions}
% \end{figure}
%

%\acknowledgements{Part of this work has been supported by grant number VH-NG-413 of the Helmholtz Association.}

%\vspace{\baselineskip}

\clearpage

\end{document}